\documentclass[traditabstract]{aa} % for the abstract without structuration 
\usepackage{graphicx}
\usepackage{natbib}
\usepackage{subfigure}
\usepackage{txfonts}
\usepackage{longtable}
\def \kms{\ifmmode{~{\rm km\,s}^{-1}}\else{~km~s$^{-1}$}\fi}
\def \vhel{\ifmmode{V_{{\rm hel}}}\else{$V_{{\rm hel}}$}\fi}
\def \vsys{\ifmmode{V_{{\rm sys}}}\else{$V_{{\rm sys}}$}\fi}
\def \degree{\ifmmode{^{\circ}}\else{$^{\circ}$}\fi}

\def \myr{\ifmmode{{\rm\ M}_\odot{\rm\ yr}^{-1}}\else{${\rm\ M}_\odot$ 
yr$^{-1}$}\fi}

\def \mdot{\ifmmode{{\rm\dot{M}}}\else{${\rm\dot{M}}$}\fi}
\def \msun{\ifmmode{{\rm\ M}_\odot}\else{${\rm\ M}_\odot$}\fi}
\def \rsun{\ifmmode{{\rm\ R}_\odot}\else{${\rm\ R}_\odot$}\fi}
\newcommand{\HA}{H$\alpha$}

\newcommand{\OIII}{[O\,{\sc iii}]\ $\lambda$5007\,\AA}

\newcommand{\SII}{[S\,{\sc ii}]\ $\lambda$6716+6731\,\AA}

\begin{document}
   \title{The post-common-envelope, binary central star of the planetary nebula Hen~2-11
   %\thanks{Based on observations made with ESO Telescopes at the La Silla Paranal Observatory under programme IDs 088.D-0573 and 090.D-0693, and the Southern African Large Telescope (SALT) under programme 2012-2-RSA-002.}
   }
   \subtitle{}

   \author{D. Jones
          \inst{1}
          \and
           H.M.J. Boffin\inst{1}
                     \and
           B. Miszalski\inst{2,3}
	\and
	R. Wesson\inst{1}
	\and
	R.L.M. Corradi\inst{4,5}
	\and
          A.A. Tyndall\inst{1,6}
                        }

   \institute{European Southern Observatory, Alonso de C\'ordova 3107, Casilla 19001, Santiago, Chile\\
              \email{djones@eso.org, hboffin@eso.org}
                       \and
             South African Astronomical Observatory, P.O. Box 9, Observatory, 7935 Cape Town, South Africa
             \and
             Southern African Large Telescope Foundation, P.O. Box 9, Observatory, 7935 Cape Town, South Africa
              \and
              Instituto de Astrof\'isica de Canarias, E-38200 La Laguna, Tenerife, Spain
              \and
              Departamento de Astrof\'isica, Universidad de La Laguna, E-38206 La Laguna, Tenerife, Spain
              \and
             Jodrell Bank Centre for  Astrophysics, School of Physics and Astronomy, University of Manchester, Manchester, M13 9PL, UK
             }

   \date{Received 4 November 2013 / Accepted 2 January 2014}

% \abstract{}{}{}{}{} 
% 5 {} token are mandatory
 
  \abstract{We present a detailed photometric study of the central star system of the planetary nebula Hen~2-11, selected for study because of its low-ionisation filaments and bipolar morphology -- traits which have been strongly linked with central star binarity.  Photometric monitoring with NTT-EFOSC2 reveals a highly irradiated, double-eclipsing, post-common-envelope system with a period of 0.609 d.  Modelling of the lightcurve indicates that the nebular progenitor is extremely hot, while the secondary in the system is probably a K-type main sequence star.  The chemical composition of the nebula is analysed, showing Hen~2-11 to be a medium-excitation non-Type \textsc{i} nebula.  A simple photoionisation model is constructed determining abundance ratios of C/O and N/O which would be consistent with the common-envelope cutting short the AGB evolution of the nebular progenitor.  
  
The detection of a post-common-envelope binary system at the heart of Hen~2-11 further strengthens the link between binary progeny and the formation of axisymmetric planetary nebulae with patterns of low-ionisation filaments, clearly demonstrating their use as morphological indicators of central star binarity.}

\titlerunning{The post-CE binary central star of PN Hen~2-11}

   \maketitle
%
%________________________________________________________________

\section{Introduction} 
\label{sec:intro}

It is now clear that common-envelope (CE) evolution represents an important channel for the formation of planetary nebulae (PNe), constituting a significant fraction of the total PN population \citep{miszalski09a}.  However, in-depth analysis of this population, as well as their influence on the host nebulae, is still hampered by the small sample and lack of detailed study for individual objects.  It is, therefore, critical to the field that further binary central stars are discovered and characterised.

Recent work has shown that PNe with post-CE, binary central stars show a penchant for low-ionisation filaments and axisymmetrical structures \citep{miszalski09b,miszalski11b}, as such PNe displaying these traits should offer the greatest chance of binary discovery.  While spectroscopic monitoring for radial velocity variables has shown promising results \citep[e.g.][]{boffin12b}, the most efficient methodology remains photometric monitoring \citep{miszalski09a,miszalski11a}.

\object{Hen 2-11} (PN G259.1+00.9; $\alpha=08^h37^m08.3^s$, $\delta=-39\degr25'08''$) is a southern PN first detected by \cite{henize67}.  The nebula was shown by \cite{gorny99} to display an elongated morphology with low-ionisation filaments.  Based on its similar appearance
in this image to two other post-CE PNe, NGC~6326 and NGC~6778
\citep{miszalski11b}, we selected it as a promising candidate for
photometric monitoring as part of ongoing efforts by our team to
discover binary central stars \citep{corradi11,miszalski11a,miszalski11b,tyndall13,miszalski13}. Here, we
report on the discovery of an eclipsing post-CE central star
displaying a strong irradiation effect that is reminiscent of
Abell 63 \citep{pollacco93} and Abell 46 \citep{pollacco94}.
The discovery of rare eclipsing central stars is a key step towards
measuring accurate, model-independent central star masses for a large
sample of PNe \citep{miszalski08}.

The paper is structured as follows.  Sect.\ \ref{sec:obs} outlines the observations and data reduction, Sect.\ \ref{sec:analysis} describes the light curve analysis and modelling, and the analysis of the stellar and nebular spectroscopy. Finally, Sect.\ \ref{sec:discussion} discusses the results. 

\begin{figure*}[]
\centering
\includegraphics[width=0.8\textwidth]{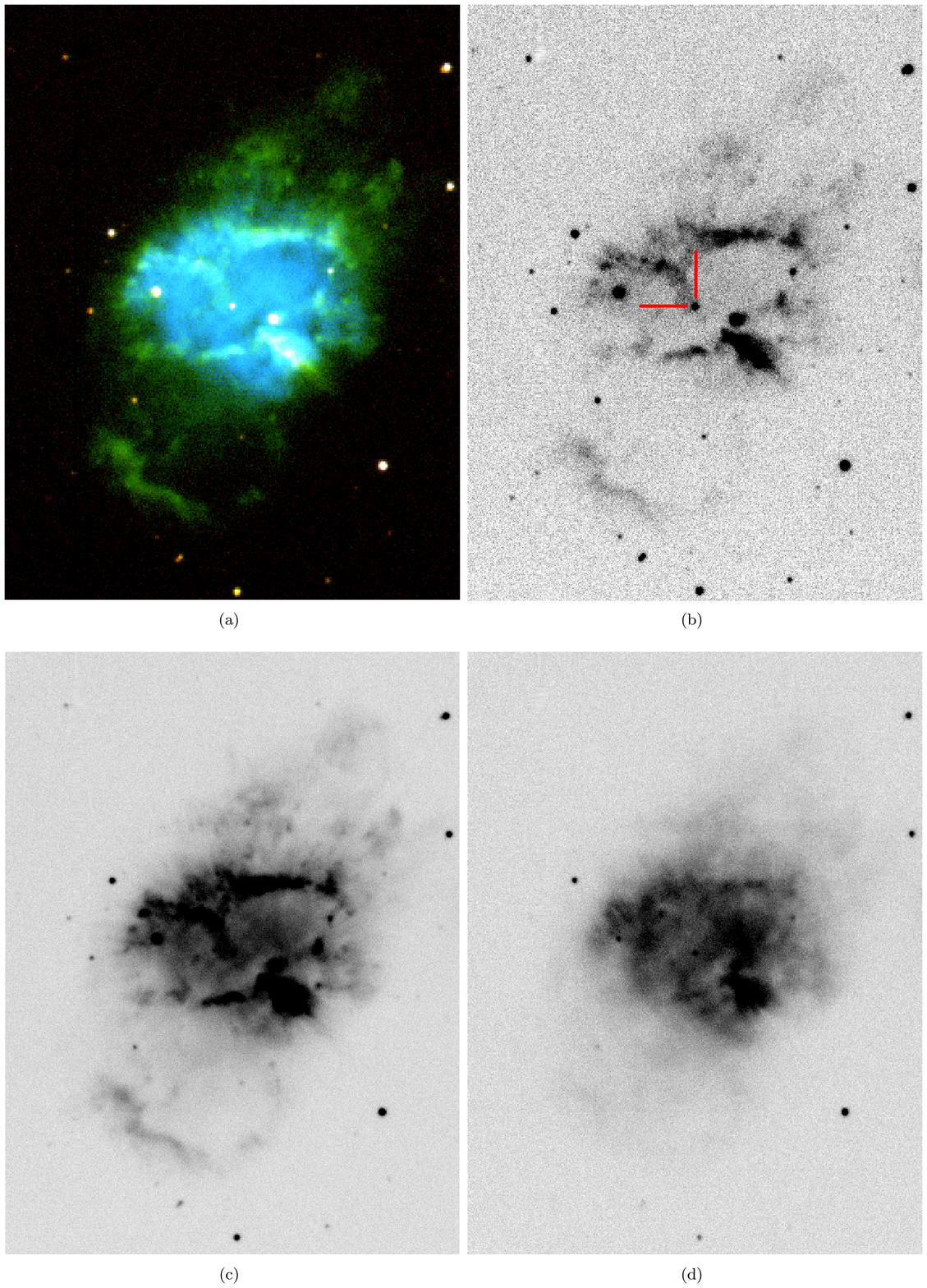}
\caption[]{NTT-EFOSC2 images of Hen~2-11. (a) Colour composite made from \SII\ (red), \HA{}+[N \textsc{ii}] (green) and \OIII\ (blue), (b) \SII\ with the central star marked, (c) \HA{}+[N~\textsc{ii}] and (d) \OIII{}. North is up, East to the left and each image measures 2\arcmin$\times$3\arcmin{}.}
\label{fig:images}
\end{figure*}

\section{Observations and data reduction}
\label{sec:obs}
\subsection{Photometry}

The central star of Hen~2-11 was monitored photometrically with EFOSC2 on the 3.6-m ESO-NTT \citep{EFOSC2a,EFOSC2b} on the nights of 27-29 February, 1-2 March 2012,  and 14, 15 and 17 January 2013.  The observations employed the i\#705 (Gunn $i$) filter and E2V CCD with a pixel scale of 0.24\arcsec{}$\times$0.24\arcsec{} pixel$^{-1}$.  A total of 200 $I$-band observations were taken with 60-s exposure time.  Further individual observations of 240-s were acquired with narrowband \OIII\ (\#687), \HA{}+[N \textsc{ii}] (\#692) and \SII\ (\#700) filters, two exposures of 20 seconds with a Bessel B filter (\#639) and five of 180-s in H$\beta$-continuum (\#743). A three-colour composite image of the nebula is presented in Fig.\ \ref{fig:images}, along with the individual images taken in the narrowband emission line filters.
 
 All data were debiased and flat-fielded using standard \textsc{starlink} routines\footnote{http://starlink.jach.hawaii.edu/}.
Photometry was extracted using \textsc{sextractor} with an aperture tailored to a diameter of 3$\times$ the seeing on each frame \citep{bertin96,jones11}, and the differential $I$-band mag of the central star measured against non-variable field stars.
An absolute scale, with an approximate precision of 0.05 mag (derived from the dispersion in detector zero points calculated for each field star), was applied to the data via the methodology described in \cite{boffin12a} using catalogue photometry from DENIS \citep{DENIS}.

\subsection{Spectroscopy}
\begin{table*}
   \centering
   \caption{Log of the SALT RSS observations.}
   \label{tab:obs}
   \begin{tabular}{llllllll}
      \hline
        Date       & JD & $\phi$ & Grating & Exptime  & $\lambda$ & $\Delta\lambda$ & Dispersion \\
        (DD/MM/YY) &  &   &    &  (s)    & (\arcsec{})          & (FWHM, \AA)     & (\AA\ pix$^{-1}$) \\
      \hline
       22/12/12 & 2456282.412835 & 0.93 & PG900 & 2350  &
3907-7004 &  6.0 & 0.98 \\
       21/03/13 & 2456373.421354 & 0.30 & PG900 & 2350 &
6300--9290 & 5.5 & 0.95 \\
       13/04/13 & 2456396.353935 & 0.94 & PG1300 & 2500 &
4082--6178 & 4.3 & 0.66 \\
      \hline
   \end{tabular}
   \tablefoot{The slit-width for all observations was 1.50\AA{}.}
  
\end{table*}

Table \ref{tab:obs} outlines
the spectroscopic observations of Hen~2-11 made with the Robert Stobie Spectrograph \citep[RSS;][]{burgh03,kobulnicky03} of the Southern African
Large Telescope \citep[SALT;][]{buckley06,odonoghue06}. The position angle (PA) of the
slit was selected to be 71.5$^\circ$, to include the brighter star to
the Southwest of the central star. Basic reductions were
applied using the PySALT\footnote{http://pysalt.salt.ac.za} package
\citep{crawford10}. Contemporaneous spectroscopic flat-fields were
taken with each PG900 spectrum and were applied to the science exposures
after each flat was divided by a 50 pixel median of itself to reduce
the amplitude of fringing on the detectors and to correct bad pixels.
Cosmic ray events were cleaned using the \textsc{lacosmic} package
\citep{vandokkum01}. Wavelength calibration of arc lamp exposures was
performed using standard \textsc{iraf}\footnote{\textsc{iraf} is distributed by the National Optical Astronomy Observatory, which is operated by the Association of Universities for Research in Astronomy (AURA) under cooperative agreement with the National Science Foundation.} tasks \textsc{identify},
\textsc{reidentify}, \textsc{fitcoords} and \textsc{transform} by
identfying the arc lines in each row and applying a geometric
transformation to the data frames. One dimensional spectra were
extracted for the central star and the nebula, keeping the same
aperture in the separate exposures.  Spectrophotometric standard stars were used to flux calibrate the
spectra in the usual fashion. Due to the moving pupil design of SALT
only a relative (not absolute) spectrophotometric solution can
be obtained.

\section{Analysis}
\label{sec:analysis}
\subsection{Morphology}

The narrowband images presented in Fig.\ \ref{fig:images} show Hen~2-11 to comprise a bright central region roughly 1\arcmin{} in diameter with an irregular patten of low-ionisation filamentary structures extending out to the Northwest and Southeast. As commented on in Sect. \ref{sec:intro}, the nebula bears a striking resemblance to two other PNe previously shown to host binary nuclei, NGC~6326 and NGC~6778 \citep{miszalski11b}.  The central region of the nebula is bright and relatively uniform in \OIII\ (Fig. \ref{fig:images}(d)), while in the light of \SII\ and \HA{}+[N \textsc{ii}] the central structure appears far more filamentary \citep[low-ionisation filaments like these are found in both NGC~6326 and NGC~6778;][]{miszalski11b}.  Away from the central region, there is little emission in \OIII{}, but the \SII\ and \HA{}+[N \textsc{ii}] images reveal bipolar extensions upto 80\arcsec{} from the central star to both the Northwest and Southeast.  A particularly bright filament is found at the tip of the Southeastern lobe, running perpendicular to the major axis of the nebula, while the Northwestern projection has a more floccular appearance. These structures may perhaps be high velocity outflows/jets similar to those seen in NGC~6778 \citep{guerrero12}, features which are hypothesised to be formed by mass transfer in a binary system \citep{miszalski13a,tocknell13}.

\subsection{Lightcurve}

\begin{figure*}[]
\centering
\includegraphics[width=\textwidth]{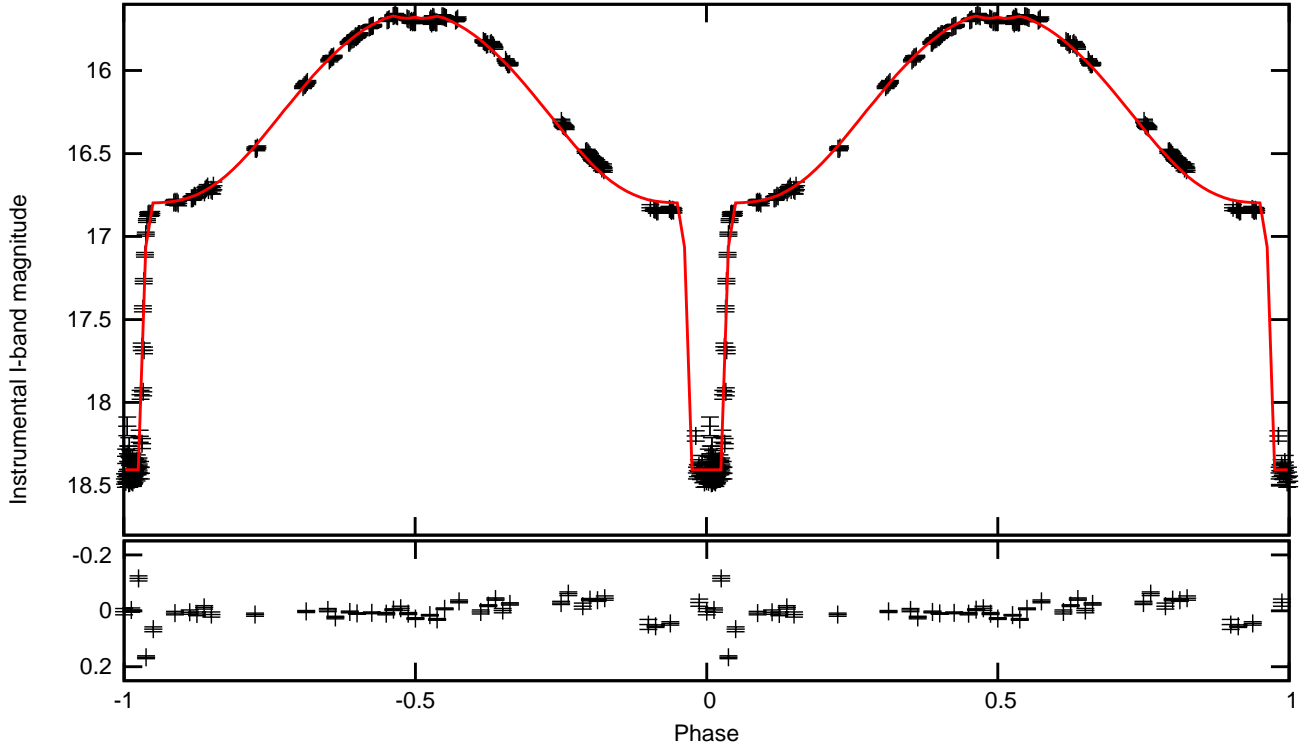}
\caption[]{Folded NTT-EFOSC2 I-band photometry of the central star of Hen~2-11 with the \textsc{nightfall} model overlaid in red (upper panel).  Binned residuals between the observed photometry and model are shown in the lower panel.  Note that the size of the points represents the photometric uncertainty and contain no estimate of the uncertainty due to variable nebular contamination (which can be significant, see text for details).}
\label{fig:phot}
\end{figure*}

A Lomb-Scargle analysis was performed in order to determine the periodicity of the variability displayed by the central star of Hen~2-11 using the \textsc{period} package of the \textsc{starlink} software suite \citep{period}.  Fig.\ \ref{fig:phot} shows the data folded on the ephemeris determined by the analysis,
$$\mathrm{min}\; I= 2\,455\,988.1617(\pm 0.0005) + 0.6093 (\pm 0.0003) E.$$

The lightcurve shows smooth sinusoidal variation of peak-to-peak amplitude $\sim$1.1 mag, with eclipses at phase 0 and 0.5.  The primary eclipse, at phase 0, is approximately 1.5 mag deep, indicating that even in the $I$-band much of the flux originates from the hot nebular progenitor.  As such, the sinusoidal variability can be attributed to the primary irradiating the near face of the secondary, and the changing projection of this face with the binary orbit (often referred to as a reflection or irradiation effect).  The strength of this reflection effect, as well as the pronounced secondary eclipse, indicate that the inclination of the system must also be close to 90\degr{}.

The system was modelled using the \textsc{nightfall} code\footnote{http://www.hs.uni-hamburg.de/DE/Ins/Per/Wichmann/Nightfall.html}.  All parameters were varied over a wide range of physical solutions, with the final model being selected for having the lowest $\chi^2$ fit.  A ``third light'' component (of 9.5\% at mean mag in the I-band) was incorporated into the model. This value is based on an estimation of the remaining nebular contamination following the background fitting and subtraction performed by \textsc{sextractor}.  The estimation was carried out by measuring the flux, on a selection of background subtracted images produced by \textsc{sextractor},  in an annulus surrounding the central star and comparing that to the flux in an aperture centred on the sky background away from the nebula, and then scaling this flux to that which would contaminate an aperture of radius 1.5$\times$ the median seeing of the observations.  Given the obviously large reflection effect in the system, detailed reflection was employed in the modelling (with 5 iterations) in order to properly treat the irradiation of the secondary by the primary.  A model atmosphere
was used for the lower temperature (secondary) component with solar metallicity and log $g$ of 4.5 \citep{kurucz93}.  The final model lightcurve is shown, along with the residuals to the binned data, in Fig.\ \ref{fig:phot} and the model parameters outlined in Tab.\ \ref{params}.

\begin{table*}
\centering
\caption{Modelled and observed binary parameters}              % title of Table
\label{params}      % is used to refer this table in the text
\centering                                      % used for centering table
\begin{tabular}{r c c}          % centered columns (4 columns)
\hline\hline                        % inserts double horizontal lines
 & Primary & Secondary \\    % table heading
\hline                                   % inserts single horizontal line
$T_\mathrm{eff}$ (K) & 140\,000$\pm$30\,000 & 4\,500$\pm$500 \\
Radius ($R_\odot$) & 0.13$\pm$0.02 & 0.68$\pm$0.03\\
Log $g$& -- & 4.5\tablefootmark{a} \\
%\hline
Inclination & \multicolumn{2}{c}{90\degr{}$\pm$0.5\degr{}}\\
$^{M_1}/_{M_2}$ & \multicolumn{2}{c}{0.91$^{+0.11}_{-0.14}$}\\\\
Period (d) & \multicolumn{2}{c}{0.6093$\pm$0.0003}\\\\
Peak-to-peak $I$-mag amplitude of irradiation effect &  \multicolumn{2}{c}{1.12$\pm$0.07}\\
Min.\ $I$-mag of sinusoidal region of lightcurve & \multicolumn{2}{c}{16.80$\pm$0.05\tablefootmark{b}}\\
Eclipse $I$-mag & \multicolumn{2}{c}{18.40$\pm$0.05\tablefootmark{c}}\\
Extinction-corrected min.\ $V$-mag of sinusoidal region & \multicolumn{2}{c}{13.82$\pm$0.08\tablefootmark{d}}\\
Extinction-corrected eclipse $V$-mag & \multicolumn{2}{c}{17.36$\pm$0.08\tablefootmark{e}}\\
\hline                                             %inserts single line
\end{tabular}
\tablefoot{
\tablefoottext{a}{A fixed parameter in the modelling}
\tablefoottext{b}{Measured value, not accounting for nebular contamination of roughly 10\% at this phase}
\tablefoottext{c}{Measured value, not accounting for nebular contamination of roughly 25\% at this phase}
\tablefoottext{d}{Corrected using the measured $c(H\beta)$=2.42$\pm$0.04, and a  representative $(V-I)_0=-0.4$ for the nebular progenitor \citep{ciardullo99}}
\tablefoottext{e}{Corrected using the measured $c(H\beta)$=2.42$\pm$0.04, and assuming that the modelled temperature is representative of the spectral type of the secondary (see text for details).}
}
\end{table*}

In general, the model fits well, with residuals of less than 0.05 mag at most phases (reasonable considering the errors plotted are solely from the photometry and do not account for the variable nebular contamination).  The worst fit is attained around primary eclipse, where the percentage contribution from the nebula is at a maximum and the data is of lowest signal-to-noise, due to the intrinsic faintness of the secondary.  The additional scatter in the data around this phase could also be due to star spots, however this is impossible to ascertain given the signal-to-noise and time coverage of our data.  In order to fully assess the validity of the model, it is important to compare the model parameters to the observations (as well as theory, where possible) and ensure the two are consistent.

The model parameters (namely temperature and radius) for the secondary indicate that it has an absolute bolometric mag, $M_\mathrm{bol}$, equal to 6.7 and is roughly of spectral type K5V \citep{boyajian12}.
Adopting a bolometric correction of $-0.66$ \citep{kaler89}, a $(V-I)_0$ colour of $1.54$ \citep{ducati01} and an extinction, $c(\mathrm{H}\beta)$, of 2.41 (determined from our SALT spectroscopy, see Sec.\ \ref{sec:salt}), and using the reddening law of \citet{howarth83}, we calculate the distance to be roughly 700pc (assuming the $I$-mag at eclipse is that of the secondary and accounting for the nebular contamination at this phase of $\sim$0.3 mag).  While this distance is a little on the low side, it is still within the error margins of the statistical distance calculated by \citet[of 890 pc]{stanghellini08} and the distance by trigonometric parallax of $\sim$780pc \citep{gutierrez99}.  We recalculated the distance using the \citet{stanghellini08} method, our extinction value, a radius of 32\arcsec and log F(H$\beta$)=$-12.14$ \citep{acker92}.  This yields 817pc, more in line with the proper-motion based distance and offering a better agreement with our distance.

\begin{figure}[]
\centering
\includegraphics[angle=270,width=\columnwidth]{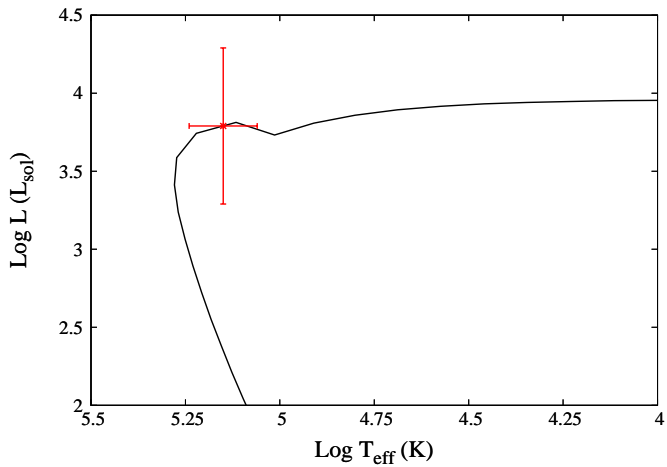}
\caption[]{Helium-burning evolutionary track for $M_i=2.5M_\odot$, $M_f=0.67M_\odot$ from \cite{vassiliadis94}, with the location of the model central star of Hen~2-11 marked (corresponding to an age of approximately 7\,000 years).}
\label{fig:evol}
\end{figure}

The spectral type of the secondary and the modelled mass ratio imply a mass for the primary of roughly 0.67 M$_\odot$.  However, determinations of spectral type based solely on modelling of the lightcurve are notoriously unreliable in these systems (see Sect.\ \ref{sec:discussion}).  It is interesting to note that in spite of the significant uncertainties, the modelled effective temperature and luminosity lie on the evolutionary track for a 0.67 M$_\odot$ remnant (formed from a 2.5 M$_\odot$ progenitor) with an age of $\sim$7\,000 years (Fig.\ \ref{fig:evol}).  This is a reasonable age for a planetary nebula, and would imply an expansion velocity, $v_{exp}$, of roughly 30--40\kms\ (given the distance to and angular size of the observed nebula), which is fairly typical for a PN with a binary central star \citep[e.g.\ ][though greater than the average for PNe in general]{jones10b}.  Collectively, this shows that our model provides a reasonable solution to the systemic parameters, in spite of the lack of radial velocity study which would be required in order to provide stronger constraints (See Sect.\ \ref{sec:discussion}).

\subsection{Stellar and nebular spectra}
\label{sec:salt}

\begin{figure}[]
\centering
\includegraphics[width=0.9\columnwidth]{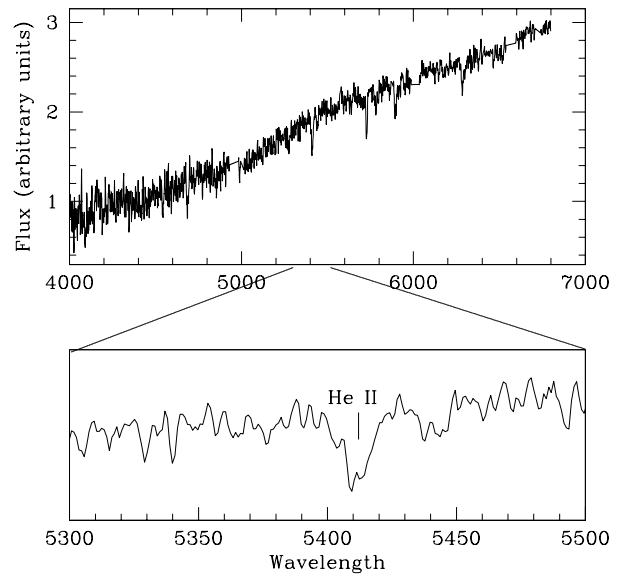}
\caption[]{Low-resolution SALT spectrum of the central star of Hen~2-11 taken at a phase of $\phi=0.93$.  The zoom displays the region around the He~\textsc{ii} $\lambda$5412\AA{} absorption feature.}
\label{fig:pg900}
\end{figure}

Figure \ref{fig:pg900} shows the low-resolution PG900 spectrum of the
central star with weak absorption from He~\textsc{ii} $\lambda$5412\AA{} detected on top of a
heavily reddened continuum, confirming it to be the ionising source of
the PN. The lack of emission lines is consistent with the irradiated
zone of the secondary facing away from us at the phase observed.

The extracted nebular spectra were combined to produce a single spectrum with wavelength coverage from $\sim$4000--9000\AA{} (shown in Fig.\ \ref{fig:nebspec}).  From the ratios of H$\alpha$, H$\gamma$ and H$\delta$ to H$\beta$, and using the Galactic extinction law of \citet{howarth83}, we determine the logarithmic extinction at H$\beta$, $c(\mathrm{H}\beta)$, to be $2.41\pm0.01$.  This is almost 0.2dex greater than that found by \citet{shaw89}, however their value is derived from emission-line photometry with appreciable contamination from [N~\textsc{ii}] $\lambda$6583\AA{} in their measurement of the H$\alpha$ flux.  Hence, we conclude that our measurement (derived using entirely non-blended spectral lines) is a more reliable determination.  Furthermore, the galactic latitude of Hen~2-11 (+0.9 degrees) means that high extinction is entirely consistent with purely foreground reddening, although some component may be internal to the object as well.

The nebular spectrum taken with the
higher resolution PG1300 grating was used to measure the heliocentric radial
velocity of the nebula, which we determine to be 22.1$\pm$2.6 km s$^{-1}$ with the
\textsc{emsao} program \citep{kurtz98}. 

\subsubsection{Plasma diagnostics}
\label{subsubsec:plasmadiags}

Using the \textsc{neat} code \citep[for full details of the analytic processes and atomic data employed, see][]{wesson12}, we measure an electron density of $570\pm90$ cm$^{-3}$ from the [S\textsc{ii}] lines, and $1800\pm1600$ cm$^{-3}$ from the [Ar \textsc{iv}] $\lambda$4711/4740\AA{} lines.  Although the uncertainty on the [Ar \textsc{iv}] density is large, the discrepancy between these two values may be due to the presence of high density clumps \citep{zhang08}; the critical densities of the [S \textsc{ii}] lines are $3.625\times10^3$ and $1.419\times10^3$ cm$^{-3}$, while those of the [Ar \textsc{iv}] lines are $1.012\times10^5$ and $1.630\times10^4$ cm$^{-3}$, and so any material at densities of $1\times10^5$ cm$^{-3}$ would emit [Ar \textsc{iv}] but not [S \textsc{ii}].

The temperatures derived from [O \textsc{iii}] and [N \textsc{ii}] are in excellent agreement at $11500\pm250$ K and $11850\pm450$ K, respectively.  The [S \textsc{iii}] $\lambda$9069/6312\AA{} line ratio implies a very high temperature, but the $\lambda$9069\AA{} line lies in a vignetted region of the spectrum and its flux is probably underestimated.  A factor of 2 increase in its measured flux would bring the [S \textsc{iii}] temperature into line with the other diagnostics.

Using the ``crossover'' method of \citet{kaler89a}, we were able to estimate the central star temperature, given the measured He~\textsc{ii}/H$\beta$ dereddened flux ratio, to be $\sim$108\,000 K -- in reasonable agreement with the temperature derived by modelling the lightcurve, 140\,000$\pm$30\,000 K\footnote{Unlike the Zanstra method, this technique does not depend on the mag of the central star, the measurement of which can be heavily skewed for binary systems where a significant proportion of the optical flux originates from the secondary.  Even for Hen~2-11, where the primary is intrinsically much brighter than the secondary, the contribution from the secondary is significant at most phases due to the strength of the irradiation effect.  This, along with the nebula's optical depth, may account for the discrepancy between our temperature and the Hydrogen Zanstra temperature of 89\,000$\pm$11\,000 K \citep{shaw89}.}.

\subsubsection{Empirical analysis}

\begin{figure}[]
\centering
\includegraphics[width=0.9\columnwidth]{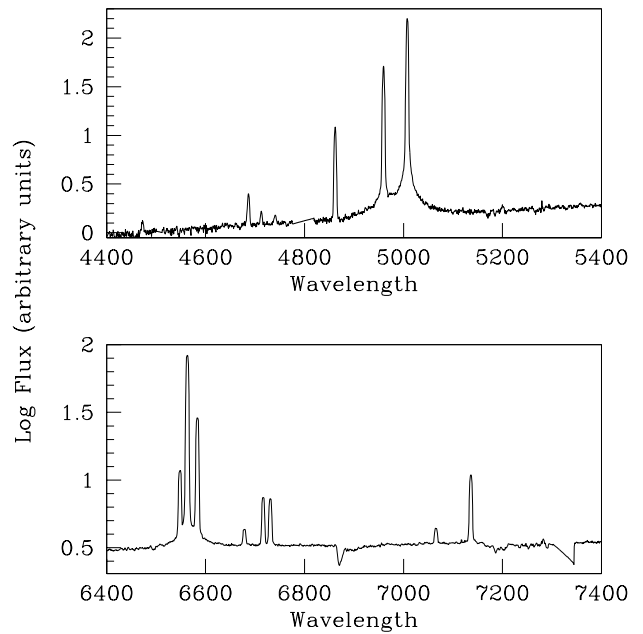}
\caption[]{A combined nebular spectrum of Hen~2-11, composited from the three exposures listed in Table \ref{tab:obs}.}
\label{fig:nebspec}
\end{figure}

We first carry out an empirical analysis using the \textsc{neat} code \citep{wesson12}, which propagates the observational uncertainties into the final quantities using a monte carlo technique.  Ionic abundances were derived using a temperature of 11\,800 K and a density of 500 cm$^{-3}$ for singly ionised species, and 11\,500 K and 1\,800 cm$^{-3}$ for more highly ionised species.  Our spectra do not extend as far bluewards as the O \textsc{ii} lines at $\lambda$3727/3729\AA{}, and the lines at $\lambda$7320/7330\AA{} lie in a gap in our spectral coverage.  Without an O$^+$/O$^{2+}$ ratio, we are unable to estimate the necessary ionisation correction factors to derive total elemental abundances for any heavy elements.  However, the presence of He$^{2+}$ (while still only comprising a relatively small amount of the total He \footnote{A greater fraction of He in He$^{2+}$ would imply that O$^{3+}$ would be the dominant ionisation stage.}) implies that O$^{2+}$ will be the dominant ionisation stage of oxygen.  
The absence of O$^+$ lines introduces a relatively small uncertainty into the O/H abundance, which we find to be $(3.3\pm0.3) \times10^{-4}$, as O$^+$/O$^{2+}$ will be low.
.The N/H abundance, on the other hand, is essentially unconstrained as only N$^+$ is observable; the fraction of N$^+$ is very small and is impossible to estimate from the current data.  We find He/H to be $0.104\pm0.007$; this indicates that Hen~2-11 is not a Type \textsc{i} PN \citep{peimbert78}. Table \ref{tab:lineflux} gives the measured and de-reddened line intensities, as well as the ionic abundances calculated using the \textsc{neat} code \citep{wesson12}, derived from
the nebular spectra shown in Fig.\ \ref{fig:nebspec}.

  \begin{table*}
   \centering
   \caption{Observed, $F\left( \lambda \right)$, and dereddened $I
\left( \lambda \right)$,  nebular emission line fluxes relative to
H$\beta=100$, and the fractional ionic abundance in that line relative
to that of Hydrogen, $\frac{X(line)}{H}$.  The final column shows the fluxes predicted by our \textsc{mocassin} photoionisation model.  The lower part of the table lists a series of emission line diagnostic ratios and their measured and modelled values.}
  \begin{tabular}{llrlrlrll}
  \hline
   $ \lambda $ & Ion & \multicolumn{2}{c}{$F \left( \lambda \right) $}&
\multicolumn{2}{c}{$I \left( \lambda \right) $}&
\multicolumn{2}{c}{$\frac{X(line)}{H}$} & $I_{model} \left( \lambda \right) $\\ \hline \hline
 3967.46+3970.07\tablefootmark{a} & [Ne~{\sc iii}]+H~{\sc i}  &   14.63& $\pm$ 1.07& 47.10& $\pm$   3.46 & - & - & $^{ \mathrm{ Ne}=29.99}_{\mathrm{ H}=15.98 }$ \\
% 3970.07 & H~{\sc i}       &  & &  & &  & & 15.98 \\
 % 4101.74 & H~{\sc i}       &    9.64& $\pm$ 0.43& 26.42& $\pm$ 1.21&-&- & 26.01 \\
  4340.47 & H~{\sc i}       &   23.64& $\pm$ 0.40& 47.77& $\pm$   0.92 &-&- & 47.00 \\
  4363.21 & [O~{\sc iii}]   &    7.09& $\pm$ 0.40& 13.91& $\pm$   0.82 & ${  4.23\times 10^{ -4}}$ & $^{+  8.89\times 10^{ -5}}_{ -1.13\times 10^{ -4}}$ & 14.08 \\
  4471.50 & He~{\sc i}      &    3.00& $\pm$ 0.39& 5.09& $\pm$   0.66 & $  0.098$ & $\pm 0.012$ & 5.41 \\
  4685.68 & He~{\sc ii}     &   14.32& $\pm$ 4.76& 13.25& $^{ +2.90}_{  -3.72}$ & ${  0.011}$ & $^{+  2.42\times 10^{ -3}}_{ -3.11\times 10^{ -3}}$ & 14.24 \\
  4711.37 & [Ar~{\sc iv}]   &    4.25& $\pm$ 0.22& 5.23& $\pm$   0.27 & ${  1.07\times 10^{ -6}}$ & $^{+  1.87\times 10^{ -7}}_{ -1.92\times 10^{ -7}}$ & 5.19 \\
  4740.17 & [Ar~{\sc iv}]   &    4.37& $\pm$ 0.86& 4.87& $^{ +0.83}_{  -1.00}$ & $  1.07\times 10^{ -6}$ & $\pm 2.03\times 10^{ -7}$ & 4.30 \\
  4861.33 & H~{\sc i}       &  100.00& $\pm$ 0.70& 100.01& $\pm$   0.00 &-&- & 100.00\\
  4958.91 & [O~{\sc iii}]   &  541.81& $\pm$ 4.87& 474.40& $\pm$   5.30 & $  3.51\times 10^{ -4}$ & $\pm  4.91\times 10^{ -5}$ & 472.30 \\
  5006.84 & [O~{\sc iii}]   & 1696.82& $\pm$ 4.82&1391.98& $\pm$   9.67 & $  3.46\times 10^{ -4}$ & $\pm  4.82\times 10^{ -5}$ & 1409.35 \\
% 5200.06 &                 &    2.93& $\pm$ 0.44& 1.85& $\pm$   0.28 \\
  5412.73 &  He~{\sc ii}               &    3.24& $\pm$   0.44&   1.54& $\pm$ 0.21 & 0.013 & $\pm$ 0.002 & 1.31 \\
  5754.60 & [N~{\sc ii}]    &    5.99& $\pm$ 0.45& 2.04& $\pm$   0.15 & ${  1.06\times 10^{ -5}}$ & $^{+  9.79\times 10^{ -7}}_{ -1.07\times 10^{ -6}}$  &1.78 \\
  5875.66 & He~{\sc i}      &   48.16& $\pm$ 0.46& 14.64& $\pm$   0.16 & ${  0.094}$ & $^{+ 7.59\times 10^{ -3}}_{ -4.87\times 10^{ -3}}$ & 14.82 \\
  6300.30 & [O~{\sc i}]     &   18.31& $\pm$ 10.94&   1.68& $^{ +0.59}_{  -0.91}$ & ${  1.84\times 10^{ -6}}$ & $^{+ 6.66\times 10^{ -7}}_{ -1.04\times 10^{ -6}}$ & 0.76 \\
  6312.10 & [S~{\sc iii}]   &   12.47& $\pm$ 11.64&   0.58& $^{ +0.27}_{  -0.50}$ & ${  6.65\times 10^{ -7}}$ & $^{+ 6.65\times 10^{ -7}}_{ -3.57\times 10^{ -7}}$ & 0.32 \\
  6363.77 & [O~{\sc i}]     &    7.43& $\pm$ 10.59&   0.07& $^{ +0.07}_{  -0.14}$ & ${  2.39\times 10^{ -7}}$ & $^{+ 2.39\times 10^{ -7}}_{ -4.73\times 10^{ -7}}$ & 0.38\\
  6548.10 & [N~{\sc ii}]    &  197.20& $\pm$ 12.59& 33.77& $\pm$   2.16 & $  1.16\times 10^{ -5}$ & $\pm  1.38\times 10^{ -6}$ & 36.44 \\
  6562.77 & H~{\sc i}       & 1680.91& $\pm$ 11.84& 284.62& $^{ +0.87}_{  -0.71}$&-&- & 283.83 \\
  6583.50 & [N~{\sc ii}]    &  549.71& $\pm$ 11.99& 91.53& $\pm$   2.13 & $  1.03\times 10^{ -5}$ & $\pm  9.80\times 10^{ -7}$ & 111.30 \\
  6678.16 & He~{\sc i}      &   25.82& $\pm$ 10.08&   2.60& $^{ +0.65}_{  -0.86}$ & ${  0.067}$ & $^{+  0.016}_{ -0.022}$ & 4.14 \\
  6716.44 & [S~{\sc ii}]    &   88.57& $\pm$ 2.06& 13.30& $\pm$   0.33 & $  4.29\times 10^{ -7}$ & $\pm  3.62\times 10^{ -8}$ &13.16 \\
  6730.82 & [S~{\sc ii}]    &   85.22& $\pm$ 2.07& 12.66& $\pm$   0.33 & $  4.30\times 10^{ -7}$ & $\pm  3.63\times 10^{ -8}$ & 12.64 \\
  7065.25 & He~{\sc i}      &   21.84& $\pm$ 2.10& 2.54& $\pm$   0.25 & $  0.102$ & $\pm 0.010$ & 2.96 \\
  7135.80 & [Ar~{\sc iii}]  &  151.84& $\pm$ 2.02& 16.84& $\pm$   0.28 & $  1.21\times 10^{ -6}$ & $\pm  1.23\times 10^{ -7}$ & 12.59 \\
  7281.35 & He~{\sc i}      &    6.09& $\pm$ 5.95& 0.13& $^{ +0.06}_{  -0.11}$ & ${  0.02}$ & $^{+  8.09\times 10^{ -3}}_{ -0.02}$ & 0.84 \\
  7751.43 & [Ar~{\sc iii}]  &   52.08& $\pm$ 2.98& 3.91& $\pm$   0.23 & $  1.17\times 10^{ -6}$ & $\pm  1.37\times 10^{ -7}$ & 3.02 \\
  %8046.71 &                 &    6.47& $\pm$ 2.82&   0.25& $^{+0.07}_{  -0.09}$ \\
  9068.60 & [S~{\sc iii}]   &  352.73& $\pm$ 4.24& 13.75& $\pm$   0.26 & $  1.42\times 10^{ -6}$ & $\pm  1.25\times 10^{ -7}$ & 8.04\\
  \hline
  4740/4711  & [Ar~{\sc iv}] & - & - & 0.93 & $^{+  0.20}_{ -0.16}$ & - & - & 0.83 \\
  6731/6717  & [S~{\sc ii}]  & - & - & 0.95 & $\pm{  0.03}$ & - & - & 0.96 \\
  (4959+5007)/4363 & [O~{\sc iii}] & - & - & 134.05 & $\pm{  7.96}$ & - & - & 133.60 \\
  (6584+6548)/5754 & [N~{\sc ii}]  & - & - & 61.94 & $\pm{  4.92}$ & - & - & 82.92 \\

\end{tabular}
     \label{tab:lineflux}
     \tablefoot{
     \tablefoottext{a}{The value measured in the spectrum is a blend of the two lines listed, while the model values are for the individual components}}
  \end{table*}

\subsubsection{Photoionisation modelling}
Given that we are unable to estimate the ionisation correction factors needed to empirically derive  heavy element abundances, we constructed a photoionisation model using \textsc{mocassin} \citep{ercolano03} to try constrain the N abundance (and further confirm the non-Type \textsc{i} nature of Hen~2-11).

From the images of the nebula (Fig. \ref{fig:images}) it is clear that the nebula is irregular and filamentary.  However, our long slit spectrum does not provide sufficient observational constraints to realistically model the nebular morphology, nor do we have any information about the surface abundances of the central star which would allow us to realistically estimate the ionising spectrum.  Our model is, therefore, highly simplified and is intended only as a first approximation, a more rigorous modelling would require spatially resolved spectroscopy of the nebula.

We modelled the nebula as a sphere with a radius of 10$^{18}$cm, estimated from our images and the distance of 700 pc derived in Section 3.2. Our model had a uniform hydrogen density, with the central star approximated by a blackbody. We then varied the nebular abundances and the central star temperature to achieve a good match to our observed line fluxes.

Our best fitting model has a stellar luminosity of 5140 L$_\odot$ and a temperature of 115\,000 K, in good agreement with the values derived in Section 3.3.1 although slightly lower than the temperature derived from the light curve modelling (but still within the uncertainty). The emission line fluxes predicted by the model are given alongside the observed values in Tab. 3.

The model reproduces most line intensities well, in particular the strengths of He~\textsc{i} and He~\textsc{ii}, and the O~\textsc{iii} nebular to auroral line ratio.  The observed [N~\textsc{ii}] temperature diagnostic ratio is lower than that predicted by the model, indicating that the actual N$^+$ temperature is higher, and the N$^+$ abundance even lower than predicted by the model.  This may indicate the presence of high-temperature low-ionisation regions of the nebula which are not captured by our simple model (see Sec. 4 for further discussion).

The total abundances used in our model are given in Tab. 4. These strengthen the indication from the empirical analysis that Hen~2-11 does not have Type \textsc{i} abundances. In particular, the nitrogen abundance in the model is very low, with N/O$\sim$0.05; Galactic Type \textsc{i} PNe have N/O$>$0.8 \citep{kingsburgh94}. While this cannot be considered a rigorous determination of the true total abundances, and therefore the true N/O ratio, it does provide a strong indication that the true N/O value will be much lower than for a typical Type \textsc{i} PN.

\begin{table}
\centering
\caption{Abundances used in the photoionisation model of Hen~2-11.}
\begin{tabular}{ll}
\hline 
Element & Abundance\\
\hline \hline
H & 1.0\\
He & 0.104\\
C & $0.8\times10^{-4}$\\
N & $3.0\times10^{-5}$\\
O & $5.5\times10^{-4}$\\
Ne & $7.5\times10^{-5}$\\
S & $1.1\times10^{-6}$\\
Ar & $1.8\times10^{-6}$\\ 
\hline
\end{tabular}
\label{tab:mocassin}
\end{table}

\section{Discussion}
\label{sec:discussion}

The central star of Hen~2-11 has been shown to be a photometric variable with a period of 0.61 d.  Modelling of the lightcurve indicates that the system is a post-CE binary with a hot primary, also the nebular progenitor, of temperature of $110-140$ kK, and a K-type main sequence secondary of temperature $\approx4500$ K.  These values, as well as the other modelled stellar parameters (size/luminosity), are roughly consistent with evolutionary models and the distances published in the literature.

It is important to point out that given the lack of radial velocity measurements, we can place only very loose constraints on the mass ratio and therefore on the mass of the primary (particularly knowing that the spectral type of the secondary, determined from its modelled temperature, is often at odds with the measured mass in these systems; see below for further discussion)  However, our model's best-fitting value of 0.67 M$_\odot$ is a fairly typical remnant mass and fits well with a similarly typical PN age (even accounting for the fairly large uncertainty in the relatively arbitrary starting time of the evolutionary tracks).  Without further observations we cannot place any stronger constraints on the system parameters, other than to say that all aspects of the model are consistent with current observations and those of similar systems.

For the modelled system, we would expect to observe radial velocity variations of around 40 \kms ($K_2 \approx 20$ \kms{} about our measured nebular heliocentric velocity of $\approx$21 \kms{}).  We, therefore, strongly encourage further spectroscopic follow-up of the system in the form of a time-resolved radial velocity study.  Given the extremely large reflection effect in the system, one would expect the secondary to show very strong C~\textsc{iii} and N~\textsc{iii} emission lines at almost all phases \citep{corradi11,miszalski11b}, thus making Hen~2-11 particularly apt for radial velocity study in spite of its intrinsic faintness.  Unfortunately, our two spectra covering these lines were taken very close to primary eclipse such that the irradiated face of the secondary was not observed, and therefore these lines were not detected. However, such a radial velocity study would allow the determination of  key system parameters such as the mass ratio and individual component masses.  The component masses are essential, not only to test the model, but also to compare the binary parameters to those of other post-CE central stars.  In general, the secondaries, in the few systems that have been the subject of detailed study, are found to display inflated radii and increased spectral temperatures with respect to those expected from their masses \citep{pollacco93,demarco08, afsar08}. As such, it would be of great interest to add further statistics in this small sample.  If this were the case for Hen~2-11, the secondary mass might be expected to be more like that of a late K- or early M-type main sequence star \citep[reducing the primary mass to possibly be more in line with that of Abell~46, another eclipsing post-CE central star displaying a large reflection effect, assuming a similar mass ratio to that determined by the lightcurve modelling;][]{pollacco94}.

Little data exists on the nebular abundances of post-CE PNe, unfortunately our data do not permit us to perform a full analysis and derive total elemental abundances.  We have derived ionic abundances where possible, and shown that Hen~2-11 is not a Peimbert Type \textsc{i} nebula.  This could be taken as an indication of the common-envelope cutting short the AGB evolution of the nebular progenitor, but this is difficult to assess without accurate C/O and N/O ratios \citep{demarco09}.  We, therefore, constructed a photoionisation model of the nebula to provide an estimate of these ratios and test the validity of this scenario.  Unfortunately, as clearly shown by the imagery in Fig. \ref{fig:images}, the nebula is far from homogenous and this is borne out in the results of our modelling.  These inhomogeneities are present throughout the nebula not only in density but also in temperature and chemistry, meaning that the derived estimates for C/O and N/O in the nebula are highly uncertain (N/O particularly so, given the discrepancy between observed and modelled temperature) meaning that the true values may indeed be greater.  However, in spite of this uncertainty, it remains clear that the nebula is both carbon and nitrogen poor, consistent with a binary partner cutting short the AGB evolution of the nebular progenitor.

The recent spate of discoveries, of short period CSPNe with main sequence secondaries \citep[this work;][]{miszalski11b,corradi11}, further highlights the discrepancy between the observed post-CE period distribution and that predicted by population synthesis models \citep[even when accounting for observational biases,][]{demarco08,rebassa08,miszalski09a}.  Knowledge of all system parameters (only possible through radial velocity study) is key in further constraining the downfalls of these population synthesis models.

Morphologically, Hen~2-11 appears remarkably similar to two other PNe proven to host post-CE central stars, NGC~6326 and NGC~6778.  However, in spite of the fact that their central binaries have shorter periods (0.372 and 0.1534 d, respectively), that of Hen~2-11 shows a much more pronounced reflection effect.  The strength of this effect depends on several factors including inclination, temperature and mass ratio.  In the case of NGC~6778 (the shortest period of the three), inclination would not seem to be an adequate explanation given its eclipsing nature and the measured inclination of the nebula \citep[$\sim$78\degr{};][]{guerrero12}.  Therefore the difference in amplitude could be attributed to a combination of NGC~6778 hosting a colder white dwarf and a difference in mass ratio between the two systems.  The difference in mass ratio is difficult to assess without detailed modelling of the central binary of NGC~6778 \citep[however it is estimated to be around 0.5;][]{guerrero12}, but a colder central star is clearly feasible given that Hen~2-11 is roughly at the hottest point of its evolution (Fig.\ \ref{fig:evol}) and the age difference between the systems implies NGC~6778 should be yet to reach that point \citep[around 2\,000 years c.f.\ 7\,000 years for Hen~2-11;][]{guerrero12}.  This hypothesis is further supported by the temperatures of the central stars, where that of Hen~2-11 is found to be $\sim$108\,000 K \citep[using our measured He~\textsc{ii}/H$\beta$ dereddened flux ratio and the relation of][ and in reasonable agreement with the primary temperature in our best fitting lightcurve model]{kaler89a}  c.f. $\sim$50\,000 K for NGC6778 \citep[derived using a similar methodology, known as the energy balance method, by][]{preite83}.

In addition to detailed study of their central stars, it is important that these objects are subjected to detailed spatio-kinematical study, in order to ascertain the three-dimensional morphology and velocity structure of the nebula, which can then be related to the evolution and parameters of the central star system \citep[particularly the inclination of nebular symmetry axis with respect to that of the binary orbital plane;][]{jones11c,jones12,tyndall12}.  To that end, we strongly encourage the acquisition of high-resolution, spatially-resolved spectroscopy (either longslit or integral field) of the nebulae Hen~2-11 and NGC~6326 \citep[NGC~6778 has already been the subject of detailed study;][]{guerrero12}.  This data would also indicate whether the structures in the Northwest and Southeastern parts of the nebula are akin to the jets/outflows seen in other binary PNe, and if so whether they were formed contemporaneously to the central region or before -- a strong indication of a phase of pre-CE mass transfer \citep{boffin12b,corradi11,miszalski13a}.

Finally, it is worthwhile to note that Hen~2-11 has been observed with the Chandra X-ray Observatory as part of the Chandra Planetary Nebulae Survey \citep[ChanPlaNS;][]{kastner12}, and, as such, represents an important test case of a close binary central star in the local sample.

\begin{acknowledgements}
We thank the anonymous referee for their comments which helped to improve the clarity of the manuscript.

Based on observations made with ESO Telescopes at the La Silla Paranal Observatory under programme IDs 088.D-0573 and 090.D-0693, and the Southern African Large Telescope (SALT) under programme 2012-2-RSA-002.  This work was co-funded under the Marie Curie Actions of the European Commission (FP7-COFUND).  This research has made use of NASA's Astrophysics Data System Bibliographic Services. This research has made use of the SIMBAD database, operated at CDS, Strasbourg, France.
\end{acknowledgements}

\bibliographystyle{aa}
\bibliography{literature}

\end{document}